\newcommand{\s}{\mbox{$\sigma $}}
\newcommand{\si}{\mbox{$\sigma _{1}$}}
\newcommand{\st}{\mbox{$\sigma _{2}$}}
\newcommand{\sth}{\mbox{$\sigma _{3}$}}

\newcommand{\e}{\mbox{$e^{ik.X(z)}$}}
\newcommand{\qe}{\mbox{$e^{iq.X(w)}$}}

\documentstyle[12pt]{article}
\newcommand{\kim}{\mbox {$ k_{1}^{\mu}$}}
\newcommand{\kom}{\mbox {$ k_{0}^{\mu}$}}
\newcommand{\ki}{\mbox {$ k_{1}$}}
\newcommand{\kt}{\mbox {$ k_{2}$}}

\newcommand{\qt}{\mbox {$ q_{2}$}}
\newcommand{\qi}{\mbox {$ q_{1}$}}

\newcommand{\qo}{\mbox {$ q_{0}$}}
\newcommand{\ko}{\mbox {$ k_{0}$}}

\newcommand{\kin}{\mbox {$ k_{1}^{\nu}$}}
\newcommand{\kon}{\mbox {$ k_{0}^{\nu}$}}
\newcommand{\ktm}{\mbox {$ k_{2}^{\mu}$}}
\newcommand{\ktn}{\mbox {$ k_{2}^{\nu}$}}

\newcommand{\li}{\mbox {$    \lambda_{1}$}}
\newcommand{\lt}{\mbox {$    \lambda_{2}$}}

\newcommand{\lpp}{\mbox {$e^{i\int _{c} \alpha (t)
k(t) \partial _{z} X(z+t) dt +ik_{0}X}$}}

\newcommand{\be}{\begin{equation}}
\newcommand{\br}{\begin{eqnarray}}
\newcommand{\ee}{\end{equation}}
\newcommand{\er}{\end{eqnarray}}

\newcommand{\gvk}{\mbox {$ e^{i\sum _{n\geq 0}k_{n}Y_{n}}$}}

\newcommand{\p}{\mbox {$ \partial$}}

\newcommand{\pp}{\mbox {$ \partial ^{2}$}}
\newcommand{\mup}{\mbox {$ \partial _{\mu}$}}
\newcommand{\nup}{\mbox {$ \partial _{\nu}$}}

\begin{document}
\title{Loop Variables and Gauge Invariant Interactions
of Massive Modes in String Theory}
\author{B. Sathiapalan\\ {\em
Physics Department}\\{\em Penn State University}\\{\em 120
Ridge View Drive}\\{\em Dunmore, PA 18512}}
\maketitle
\begin{abstract}
The loop variable approach used earlier to obtain free equations of
motion for the massive modes of the open string, is generalized to
include interaction terms.  These terms, which are polynomial,
involve only modes of strictly lower mass.  Considerations based
on operator product expansions suggest that these equations are
particular truncations of the full string equations.  The method
involves broadening the loop to a band of finite thickness that
describes all the different interacting strings.  Interestingly,
in terms of these variables, the theory appears non-interacting.
\end{abstract}
\newpage
\section{Introduction}

A physically compelling symmetry principle would greatly advance our
understanding of string theory.  In the first quantized formalism,
reparametrization invariance and Weyl invariance have been the
guiding principles.  This is particularly transparent in the Polyakov
approach \cite{poly}.  Via the BRST device this has been carried over
into second quantized string field theory
\cite{s,sz,w2,lpp,z}.  A space-time interpretation for these symmetries
would thus be very useful.

In \cite{bslv,bslv1}, following the renormalization group approach
\cite{l,cc,as,ft}, the gauge invariant free equations
and transformation laws of
string fields were written down.  The basic ingredient
is a `loop variable', which is essentially a collection
of all the vertex operators of first quantized string theory.
In this approach the
gauge transformations have a simple form. In \cite{bslv}
there was also some speculation about
a possible space-time interpretation.

In this paper we would like to report on  some progress in the study of
(gauge invariant) interactions in this `loop variable' approach. We will
deviate a little from the renormalization group approach that has been
used so far [11-34].  Starting from the free equations of motion
and the free
gauge transformation laws of \cite{bslv,bslv1}, we will make an educated
``guess'' as to the form of the interacting equations and
transformation laws.
The result will be an interacting equation of motion, the form of which is
such that the invariance of these equations under the
modified gauge transformation follows naturally from the
invariance of the free equations.  This is reminiscent of the usual
field theory trick of obtaining interactions by covariantising derivatives.
However, in detail, our approach is quite different. \footnote{We will
be dealing, in any case, with strings that carry no Yang-Mills or
U(1) quantum numbers.}  In fact, the interacting equations look
like the free equations of a string that has been broadened into a `band'.
This is an intriguing result and
one can speculate that in some approximation
an {\em interacting} string behaves like a {\em free} membrane.

We should, however, emphasize that we do not obtain the
full set of interactions
of string theory.  The equation for
any particular mode contains only terms involving modes of strictly
lower mass.  In effect, we are picking the `naive product'-term
\footnote{By `naive product', we mean the term that involves no
contractions.
This will be made precise in the last section.}
in the
infinite series of terms contained in the o
perator product expansion (OPE)
of two (or more) vertex operators.
Thus each equation is a truncation of the full equation, where higher
mass fields have been set to zero and in addition higher order terms
involving fields of the same mass are also set to zero.  Only terms that
have the same oscillator number ($\sum _{n} n a^{\dag}_{n}a_{n}$)
are present in any equation.  This truncation is consistent, in the sense
that the equations are gauge invariant, with the same number of
gauge parameters as the free theory.
Hopefully, it should be possible to
generalize the construction to include the full set of interactions.

This paper is organized as follows: Section II describes how
interacting equations and transformation laws are obtained from the
free ones.  Section III makes contact with string theory
by describing how the field content is restricted to
agree with that of string theory and how massive
equations are obtained from massless ones, as in the loop variable approach.
Section IV contains some conclusions
and also explains the connection with the OPE of vertex operators
and the nature of the truncation that is being performed
on the full equation.

\newpage
\section{Interactions and Loop Variables}
In this section we describe how one obtains interacting equations
starting from free ones.  We refer the reader to \cite{bslv,bslv1}
for a description of how one obtains the free equations of motion.

The loop variable is defined by
\be     \label{2.1}
\lpp
\ee
\be     \label{2.2}
= \, \gvk
\ee
where
\[
Y_{n} =
\frac{\p ^{n} X}{(n-1)!} +
\alpha _{1}\frac{\p ^{n+1} X}{n!} + ...
\alpha _{m}\frac{\p ^{n+m} X}{(n+m-1)!} +...
\]
and
\be     \label{2.3}
\alpha (t)= \alpha _{0}+
\frac{\alpha _{1}}{t} +
\frac{\alpha _{2}}{t^{2}} +...
\ee
The space-time fields are obtained from a string field $\Psi [k_{n}]$
as follows:($[dk_{n}] \equiv d \ki d\kt ... dk_{n} ...$ and
$\Psi [k_{n}] \equiv \Psi (\ko , \ki , \kt , ...k_{n} ...) $)

\be     \label{2.4}
\int [dk_{n}] \Psi [k_{n}] \, \equiv \, \phi (\ko )
\ee
is the tachyon,
\be     \label{2.5}
\int [dk_{n}] \kim \Psi [k_{n}] \, \equiv \, A^{\mu} (\ko )
\ee
is the vector,
\be     \label{2.6}
\int [dk_{n}] \ktm \Psi [k_{n}] \, \equiv \, S^{\mu} (\ko )
\ee
and
\be     \label{2.7}
\int [dk_{n}] \kim \kin \Psi [k_{n}] \, \equiv \, S^{\mu \nu} (\ko )
\ee
are the
(auxiliary) spin-1 and spin-2 fields.

In order to define an interacting theory, we introduce an additional
coordinate `$\sigma $',($0 \leq \sigma \leq 1$),
to parametrize different space-time fields as follows:
We make $k(t)$ a function of $\sigma $ as well, i.e. $k(t, \s )$.  Thus
$k_{i}$ in eq.(\ref{2.2}) becomes $k_{i}(\s )$.  Thus $\Psi $ also becomes
an implicit function of $\s $: $ \Psi  [k _{n}(\s )]$. We retain
eqns. (\ref{2.4}),(\ref{2.5}) and (\ref{2.6}) and modify (\ref{2.7})
to
\be     \label{2.8}
\int [{\cal D}k_{n}] \kim (\s _{1}) \kin (\s _{2})
\Psi [k_{n}] \, = \, \delta (\s _{1}- \s _{2}) S^{\mu \nu}
+A^{\mu}(\ko )   A^{\nu}(\qo )
\ee
We have replaced $dk_{n}$ by ${\cal D} k_{n}$ to denote a functional
integral.  Eqn.(\ref{2.8}) has the following interpretation:
When $\s _{1}$ coincides with $\s _{2}$ we get (a higher excitation of) the
same string, but otherwise we get two different strings at the same
space-time point - which is an interaction.  $\ko (\s )$ stands
for the momentum of a particular string.  Thus, in (\ref{2.8}),
we have called $\ko (\s _{1})$ and $\ko (\s _{2})$,
$\ko$ and $\qo$ respectively.

One can thus imagine that $\Psi [k _{n}(\s )]$ defines a field theory,
where the fields are $k_{i} (\s ) $.  (\ref{2.8}) then describes
the two point correlation function of $\ki (\s )$:
\be     \label{2.9}
<\kim (\s _{1} ) \kin (\s _{2} ) >= \delta (\s _{1} - \s_{2}) S^{\mu \nu}
+<\kim (\s _{1} )>< \kin (\s _{2})>
\ee
where, by (\ref{2.5}),
\be     \label{2.10}
<\kim (\s _{1} )> \, =\, A^{\mu } (\ko )
\ee
\be     \label{2.11}
<\kim (\s _{2} )> \, =\, A^{\mu } (\qo )
\ee
One can generalize this to three point correlation functions:

Thus
\[
\int [dk_{n}] \kim (\s _{1})\kin (\s _{2})\ki ^{\rho} (\s _{3})
\, = \,
\delta (\s _{1} - \s _{2})\delta (\s _{2} - \s _{3})
\phi _{111}^{\mu \nu \rho} +
\]
\be     \label{2.12}
[\delta (\s _{1} - \s _{2})S^{\mu \nu} A^{\rho} +
\delta (\s _{1} - \s _{3}) S^{\mu \rho} A^{\nu} +
\delta (\s _{2} - \s _{3}) S^{\rho \nu} A^{\mu}] + A^{\mu} A^{\nu}A^{\rho}
\ee
If we integrate over $\s _{1}, \s _{2},\s _{3}$, RHS of (\ref{2.12})
becomes
\be     \label{2.12.1}
\phi _{111}^{\mu \nu \rho} + S^{\mu \nu } A^{\rho} +S^{\mu \rho}A^{\nu}
+ S^{\nu \rho}A^{\mu} + A^{\mu}A^{\nu}A^{\rho}
\ee
(\ref{2.12}) describes a higher string excitation along with a
quadratic and a cubic interaction.  The generalization to N-point
correlation functions is obvious. One can also define correlators
involving $\ko (\s )$.  Thus
\be     \label{2.13}
\int [{\cal D}k_{n} (\s )]\kom (\s _{1}) \Psi [k_{n}] \, =\,
\kom \phi (\ko )
\ee
\be     \label{2.14}
\int [{\cal D}k_{n} (\s )]\kom (\s _{1}) \kin (\s _{2})\Psi [k_{n}]
= \, \delta (\s _{1} -\s _{2}) \kom A^{\nu}(\ko ) +
\kom \phi (\ko ) A^{\nu} (\qo )
\ee
which can be rewritten as $e^{-\phi} \p ^{\mu} e^{\phi} A^{\nu}$.
\[
\int [{\cal D}k_{n} (\s )]\kom (\s _{1}) \kin (\s _{2})
\ki ^{\rho} (\s _{3})\Psi [k_{n}]
\]
\[
=\,
\delta (\s _{1} - \s _{2})\delta (\s _{2} - \s _{3})
\kom S^{\nu \rho } +
\delta (\s _{2} - \s _{3})\kom \phi (\ko ) S^{\nu \rho} +
\delta (\s _{1} - \s _{2})\kom A^{\nu}(\ko ) A^{\rho}(p_{0}) +
\]
\be     \label{2.15}
\delta (\s _{1} - \s _{3})p_{0}^{\mu} A^{\nu}(p_{0}) A^{\rho}(\qo )+
\kom \phi (\ko )A^{\nu}(\qo ) A^{\rho}(p_{0})
\ee
If we integrate over $\s _{1},\s _{2}, \s _{3}$ eq.(\ref{2.15}) can be
summarized as
\be     \label{2.16}
e^{-\phi}\mup e^{\phi} (S^{\nu \rho} + A^{\nu}A^{\rho})
\ee
For convenience, henceforth we will set $\phi =0$.

Using the above, the prescription for obtaining
interacting equations is very simple:
Take the free equation and replace $k _{i}$ by $\int d \s k_{i} (\s )$
and use eqns(\ref{2.8}),(\ref{2.12}),(\ref{2.13})-(\ref{2.15}).  In terms
of fields all we are doing is replacing $S^{\mu \nu}$ by
$S^{\mu \nu} + A^{\mu}A^{\nu}$ and $\phi_{111}^{\mu \nu \rho}$
by eqn.(\ref{2.12.1}).
To obtain the $\phi$ dependence, simply replace the derivative $\mup $
by $e^{-\phi}\mup e^{\phi}$.

Now let us apply this prescription to obtain interacting equations
for $S^{\mu}$ and $S^{\mu \nu}$.  The free equation for $S^{\mu}$ is
(\cite{bslv}):
\be     \label{2.17}
\ko ^{2} \ktm - \ko . \ki \kim + \kom \ki .\ki - \kom \ko . \kt \, = \, 0.
\ee
This becomes the interacting equation:
\[
\int d\s _{1} d\s _{2} d\s _{3}\{ \kon (\si ) \ko _{\nu}(\st ) \ktm (\sth )
- \ko _{\nu}(\si ) \kin (\st ) \kim (\sth ) +
\]
\be     \label{2.17.1}
\kom (\si ) \kin (\st )
\ki _{\nu} (\sth ) - \kom (\si ) \kon (\st ) \kt _{\nu} (\sth )\} \, =\, 0.
\ee
which becomes, in terms of space-time fields:
\be     \label{2.18}
- \pp S_{\mu} -i \nup (S_{\mu}^{ \nu} + A_{\mu}A^{\nu}) +
i \mup (S^{\nu}_{\nu}
+A.A ) + \mup \nup S^{\nu}  \,=\, 0.
\ee
Similarly the free equation for
$S^{\mu \nu}$ is
\be     \label{2.19}
-\ko ^{2} \kim \kin + \ki .\ko (\kim \kon + \kin \kom ) -
\ki . \ki \kom \kon \, = \, 0.
\ee
and it becomes the interacting equation
\be     \label{2.20}
\pp (S^{\mu \nu}+ A^{\mu}A^{\nu}) + \p _{\rho} \p ^{(\nu}S^{\mu ) \rho}
+ \p _{\rho} \p ^{(\nu}(A^{\mu )} A^{\rho})
-\mup \nup (S^{\rho}_{\rho} + A^{\rho}A_{\rho}) \, = \, 0.
\ee
Equations (\ref{2.18}) and (\ref{2.20}) describe only cubic interactions
of the type $S AA$.  Equations for higher
spin fields such as $\phi _{111}^{\mu \nu \rho}$ will describe
cubic as well as quartic interactions of the type $\phi SA$ and $\phi AAA$.
However, as we shall see in the last section interactions of this type
are only a small subset of the interactions actually present in string
theory.

Eqns.(\ref{2.18}) and (\ref{2.20}) describe massless modes.  To make
contact with string theory one has to introduce masses by a process
of dimensional reduction and impose constraints on the $k_{i}$
\cite{bslv,bslv1}. We will describe this in the next section.

Let us turn, now, to the gauge transformation law.  For the free theory,
gauge transformations are succinctly described by the equation
\be     \label{2.24}
k(t) \rightarrow k(t) \lambda (t)
\ee
In terms of modes,
\be     \label{2.25}
k_{n} \rightarrow k_{n} + \lambda _{1} k_{n-1} + \lambda _{2} k_{n-2}
+ ...+ \lambda _{n} \ko
\ee

This is easily seen to leave (\ref{2.17}) and (\ref{2.19}) invariant.
To translate this into a law for spacetime fields we follow the
prescription described in \cite{bslv,bslv1}, viz., do as in
in equations (\ref{2.4})-(\ref{2.7}),
but with $\Psi$ taken to be a function of the gauge parameter
$\lambda (t)$ also.  Thus,
\be     \label{2.26}
\int [dk_{n}] [d\lambda _{n}] \li \Psi [k_{n},\lambda _{n}]
\, \equiv \, \Lambda _{1}
\ee
and
\be     \label{2.27}
\int [dk_{n}] [d\lambda _{n}] \li \kim \Psi [k_{n},\lambda _{n}]
\, \equiv \, \Lambda _{2}^{\mu}
\ee
\be     \label{2.28}
\int [dk_{n}] [d\lambda _{n}] \lt [k_{n},\lambda _{n}]
\, \equiv \, \Lambda _{2}
\ee
are some of the gauge parameters that enter the gauge transformation
laws of the space-time fields $A^{\mu} , S^{\mu} , S^{\mu \nu}$.  Using
(\ref{2.25}) and (\ref{2.26}) - (\ref{2.28}) one finds:
\be
A_{\mu} \rightarrow A_{\mu} + \mup \Lambda _{1}
\ee
\be     \label{2.28.5}
S_{\mu} \rightarrow S_{\mu} + \mup \Lambda _{2} + \Lambda _{2\mu}
\ee
\be     \label{2.29}
S_{\mu \nu}\rightarrow S_{\mu \nu} +
\mup \Lambda _{2 \nu} +\nup \Lambda _{2 \mu}
\ee

Now we have to generalize  (\ref{2.24}) and (\ref{2.25}) such that it is
a symmetry of the interacting equations. This is achieved by the following
gauge transformation:

\be     \label{2.30}
\int d \s k(t, \s )
\rightarrow \int d \s k(t, \s ) \int d \si \lambda (t, \si )
\ee
The interacting equations were obtained from the free ones by the
prescription $k(t) \rightarrow \int d \s k(t, \s )$. Thus the
invariance of the free equations under (\ref{2.24}) guarantees
that the interacting equations are invariant under (\ref{2.30}).
Of course, one has to choose the gauge transformation laws
for a space-time field so that it
is consistent with the already
fixed gauge transformation laws of the lower mass fields.  This is always
possible since one can use (\ref{2.30})
to define the gauge transformation law
of that field.  We illustrate this below.

We set $\phi =0$ for convenience.  Consider the gauge transformations
of the various $k_{i}$'s and their products.  We start with $\ki$:
\[
\delta [ \int d \s \kim ( \s ) ] \, = \, \delta A^{\mu} \,
\]
\be     \label{2.31}
=\, \int d \si \kom (\si ) \int d \st \li ( \st ) \, =\, \mup \Lambda _{1}
\ee
which is unchanged from the free case. \footnote{If $\phi \neq 0$, we
we would get $e^{-\phi}\mup e^{\phi}\Lambda _{1}$.}
\[
\delta [ \int d \s \ktm (\s ) ] = \delta S^{\mu}
\]
\[
=\, \int d \si d\st [ \kim (\si ) \li (\st ) + \kom (\si ) \lt (\st )]
\]
\[
=\, \int d\si d\st [ \delta (\si -\st ) \Lambda _{2} ^{\mu} + A^{\mu}
\Lambda _{1} + \delta (\si -\st ) \p ^{\mu}\Lambda _{2}]
\]
Thus
\be     \label{2.32}
\delta S^{\mu} \, =\, \Lambda _{2} ^{\mu}+ \p ^{\mu} \Lambda _{2} +
A^{\mu}
\Lambda _{1}
\ee
And,
\[
\delta [ \int d\si d\st \kim (\si ) \kin (\st )] \, =\, \delta [
S^{\mu \nu} + A^{\mu} A^{\nu}]
\]
\[
=\, \int d\si d\st d\sth \ko ^{(\mu}(\si ) \ki ^{\nu )} (\st )
\lambda (\sth )
\]
\[
=\, \int d\si d\st d\sth [ \delta (\si -\st ) \delta (\st - \sth )
\p ^{(\mu } \Lambda _{2} ^{\nu )} +\delta (\si - \st ) \p ^{(\mu}
A^{\nu )}\Lambda _{1} +
\]
\[
\delta (\si - \sth ) \p ^{(\mu} \Lambda _{1} A^{\nu )}]
\]
\be     \label{2.33}
=\, \p ^{(\mu } \Lambda _{2}^{\nu )} +
A^{(\mu} \p ^{\nu )} \Lambda _{1}
+ \p ^{(\mu} A^{\nu )}\Lambda _{1}
\ee

If we use (\ref{2.32}) and (\ref{2.33}) as the gauge transformation laws,
we are guaranteed that the equations will be invariant.  Using
(\ref{2.31}) and(\ref{2.33}) we find that
\be     \label{2.34}
\delta S^{\mu \nu} \, =\,
\p ^{(\mu } \Lambda _{2}^{\nu )} + \p ^{(\mu} A^{\nu )}\Lambda _{1}
\ee
Thus (\ref{2.31}),(\ref{2.32}) and (\ref{2.34}) are the gauge transformation
laws that leave the (non linear) equations invariant.

The substitutions that take one from the free equations and free
gauge transformations to
the interacting equations and gauge transformations can be summarized
by the following rules:
\[
S^{\mu \nu} \rightarrow S^{\mu \nu} + A^{\mu}A^{\nu}  ,
\]
\[
S^{\mu} \rightarrow S^{\mu} \, \, ; \, \, A^{\mu}\rightarrow A^{\mu} ,
\]
\be     \label{2.35}
\Lambda _{2} ^{\mu} \rightarrow \Lambda _{2}^{\mu} +
\Lambda _{1} A^{\mu}
\ee
It can easily be checked that the gauge transformation laws
(\ref{2.32}) and (\ref{2.33})
are obtained  from (\ref{2.28.5}) and (\ref{2.29}) by these substitutions.
It is very
easy to generalize (\ref{2.31})-(\ref{2.34}) to the third level:

\be     \label{2.36}
\int [dk_{n}] \kim \kin \ki ^{\rho} \Psi [k_{n}] \equiv
\phi _{111}^{\mu \nu \rho}
\ee
\be     \label{2.37}
\int [dk_{n}] \kt ^{(\mu} \ki ^{\nu )}  \Psi [k_{n}] \equiv
\phi _{(12)}^{\mu \nu }
\ee
\be     \label{2.39}
\int [dk_{n}] \kt ^{[\mu} \ki ^{\nu ]}  \Psi [k_{n}] \equiv
\phi _{[12]}^{\mu \nu }
\ee
\be     \label{2.40}
\int [dk_{n}] k _{3} ^{\mu}   \Psi [k_{n}] \equiv
\phi _{3}^{\mu }
\ee
The interacting version of (\ref{2.36}) was already given in
eq.(\ref{2.12.1}).  Similar calculations can easily be done
for the rest of the fields. One can then use (\ref{2.30}) to obtain the
gauge transformation law for expression (\ref{2.12.1}), whence,
using (\ref{2.31}) and (\ref{2.34}),
one deduces the gauge transformation law for
$\phi _{111}^{\mu \nu \rho}$.  Since there is nothing conceptually
new we do not give the results here.  Instead we will work out the massive
case in the next section.

In this section we have described a simple method of obtaining, starting
from the free equations for (massless) higher spin fields (described in
\cite{bslv,bslv1}), interacting equations and the corresponding
gauge transformation laws.  In the next section we make contact with
string theory following the procedure of \cite{bslv,bslv1}.

\newpage
\section{Massive Equations}
\setcounter{equation}{0}
In \cite{bslv,bslv1} a two step procedure was described for obtaining
gauge invariant massive equations starting from the massless ones, and
then for reducing the number of fields to agree with that of the
usual BRST formalism for string theory \cite{sz}.  We will apply this
procedure to the interacting theory described in Sec II and show how
one obtains equations for massive modes (that interact with other
massive and massless modes), with the field content at each mass level
being exactly that of string theory.  Furthermore, these equations
have all the gauge invariances that the free equations have. It is
therefore very likely that these equations are a truncated version
of the full tree level equations of string theory.  Consideration
of OPE of vertex operators (Sec IV) supports this view.

To obtain massive equations from massless ones, we perform a dimensional
reduction.  We split the generalized momentum $k^{\mu}(t); \, \mu = 1,...,D$
into $(k^{\mu}(t), q(t)); \mu \, = 1,...,D-1$.  We then set $\qo = \, mass $.
Thus, for the tachyon  $\qo ^{2} = -2$, for the massless vector
$\qo ^{2}=0$, for the first mass level $\qo ^{2} =2$ and so on.  We will
also set $\int [dk_{n}][dq_{n}]\qi \Psi [k_{n},q_{n}] =0$.
The necessity for this was demonstrated in \cite{sz}.  Thus at the
massless level we just have a vector, $k_{1}^{\mu}$.  At the next
level there are more fields.

\subsection{Spin-2}

There are two fields $\kim \kin $ and $\ktm$.  Equation (\ref{2.17})
splits into two equations ($\qo ^{2}=2$).
\be     \label{3.1}
(\ko ^{2} + \qo ^{2})\ktm - (\ko .\ki +\qo .\qi ) \kim +
\kom (\ki .\ki +
\qi \qi ) - \kom (\ko . \kt + \qo \qt ) \, = \, 0.
\ee
\be     \label{3.2}
(\ko ^{2} + \qo ^{2})\qt - (\ko .\ki +\qo .\qi ) \qi +
\qo (\ki .\ki +
\qi \qi ) - \qo (\ko . \kt + \qo \qt ) \, = \, 0.
\ee
Equation (\ref{2.19}) can also be split up into three equations involving
$\kim \kin$,$ \kim \qi $ and $\qi \qi$.  Furthermore we will make the
identifications
\[
\int [dk_{n}] [dq_{n}]\qi \kim \Psi [k_{n},q_{n}] \, =
\int [dk_{n}] [dq_{n}]\ktm \qo \Psi [k_{n}]
\, =\, S^{\mu} \qo
\]
and
\be     \label{3.3}
\int [dk_{n}] [dq_{n}]\qi \qi \Psi [k_{n}]
\, = \, \int [dk_{n}][dq_{n}] \qt \qo \Psi [k_{n}]
\, =\, \eta \qo
\ee
In all the equations we can use (\ref{3.3}) to eliminate $\qi$.  The
equations for $\qi \kim $ and $\qi \qi $ then reduce to (\ref{3.1})
and (\ref{3.2}).  The equation for $\kim \kin $ becomes:
\be     \label{3.4}
-(\ko ^{2} +\qo ^{2})\kim \kin + \ki . \ko (\kim \kon + \kin \kom )
+ \qo ^{2}(\ktm \kon + \ktn \kom ) - (\ki .\ki + \qt \qo ) \kom \kon
\, = \,0.
\ee

Thus we end up with three fields, $S^{\mu \nu}, \, S^{\mu}, $ and
$\eta $ - which is exactly the field content (including
auxiliary fields) at the
second mass level in string theory \cite{sz}.

We also have to eliminate $\qi$ from the gauge transformations.

Thus in
\be     \label{3.5}
\qt \rightarrow \qt + \lt \qo + \li \qi
\ee
we will set
\be     \label{3.6}
\qi \li \, = \, \lt \qo
\ee
so that
\be     \label{3.7}
\qt \rightarrow \qt + 2 \lt \qo
\ee
One can also check, then, that
\be     \label{3.8}
\qi \qi \rightarrow \qi \qi + 2 \li \qi \qo \, = \, \qi \qi +
2 \lt \qo ^{2}
\ee
Comparing (\ref{3.7}) and (\ref{3.8}) we see that (\ref{3.3})
is consistent.

Now that we have the correct field content and gauge transformations for
a free massive spin-2, we can go ahead and introduce interactions
in the manner described in Sec II.  Note that first we have to get
rid of all $\qi$-dependence in the free equations by replacing
$\qi \kim$ and $\qi \qi$ using (\ref{3.3}),
and only after that can we introduce interactions.  The net result is that
$S^{\mu \nu}$ is replaced by $S^{\mu \nu }+ A^{\mu} A^{\nu}$, just
as in Sec II.  The gauge transformations are also the same as before.

For $\eta$ we have:
\[
\delta [ \int d\s \qt (\s )]\, =\, \delta \eta \, =\, 2 \int d\si
\lt (\si ) \int d \st \qo (\st )
\]
\be     \label{3.9}
\Rightarrow  \delta \eta \, = \, 2 \qo \Lambda _{2}
\ee
Equation (\ref{3.1}) becomes, in the interacting case ($\qo ^{2}=2$):
\be     \label{3.10}
-\pp S^{\mu} -i\nup (S^{\nu \mu} + A^{\nu}A^{\mu}) + i \p ^{\mu}
(S^{\nu}_{\nu} +A^{\nu}A_{\nu})-i\p ^{\mu} \nup S^{\nu} \, =\, 0.
\ee
(\ref{3.2}) becomes:
\be     \label{3.11}
-\pp \eta - 2 i \nup S^{\nu} \qo +\qo(S^{\nu}_{\nu} +A^{\nu}A_{\nu})\, =\,0.
\ee
and (\ref{3.4}) becomes:
\[
(-\pp + \qo ^{2}) S^{\mu \nu} -[\p _{\rho} \p ^{\nu} (S^{\mu \rho} +
A^{\mu}A^{\rho}) + (\mu \leftrightarrow \nu )] +
\]
\be     \label{3.12}
\qo ^{2} \p ^{(\nu  } S^{\mu )} + \p ^{\mu} \p ^{\nu} (S^{\rho}_{\rho}
+A^{\rho}A_{\rho} + \eta \qo ) \,= \, 0.
\ee

Thus we have a set of equations for a
massive spin-2 system interacting with a massless vector $A^{\mu}$.
Note that (\ref{3.10}) and (\ref{3.11}) are unchanged from the massless case,
while (\ref{3.12}) is changed.
The gauge transformations are given by (\ref{2.31}), (\ref{2.32}),
(\ref{2.34}) and (\ref{3.9}).  We reproduce them here for convenience.
($\qo ^{2}=2$)
\[
\delta A ^{\mu} = \p ^{\mu} \Lambda _{1}
\]
\[
\delta S ^{\mu} = \p ^{\mu} \Lambda _{2} + A^{\mu}\Lambda _{1} +
\Lambda _{2}^{\mu}
\]
\[
\delta S^{\mu \nu} \, =\, \p ^{(\mu } \Lambda _{2}^{\nu )}
+ \Lambda _{1} \p ^{(\mu } A^{\nu )}
\]
\be     \label{3.13}
\delta \eta \, =\, 2 \qo \Lambda _{2}
\ee

\subsection{Spin-3}

We now turn to the spin-3 case.  We only give an outline.  The equations
for the free theory are given in \cite{bslv}.  The procedure for obtaining
the interacting theory was described in Sec II.  But before we apply this
prescription, we have to first perform the dimensional reduction
and impose constraints similar to (\ref{3.3}) in order to obtain the
correct field content.  In order to do this we have to eliminate (terms
involving) $\qi$ from the equations.  The various fields are:
\[
\kim \kin \ki ^{\rho}, \, \qi \kim \kin , \, \ktm \kin ,\,
\qi \qi \kim , \, \qt \kim ,
\]
\be     \label{3.14}
\qi \ktm , \, k_{3}^{\mu} ,
\qi \qi \qi , \qi \qt , \, and \, q_{3}
\ee
In making identifications between fields, it is crucial that they do not
involve derivatives, since this could introduce higher derivatives in the
kinetic terms.  We start with the fields that have two Lorentz indices,
$\qi \kim \kin $ and $\ktm \kin$.  Let us compare their transformation
laws:
\be     \label{3.15}
\delta [2\qi \kim \kin ] \, = \, 2 \li \qo \kim \kin +
2 \li \qi \ko ^{(\mu}\ki ^{\nu )}
\ee
\be     \label{3.16}
\delta [\qo \kt ^{(\mu} \ki^{\nu )} ] \, = \, 2 \li \qo \kim \kin +
\qo \lt \ko ^{(\mu}\ki ^{\nu )} + \qo \li \ko ^{(\mu}\kt ^{\nu )}
\ee
If we want the terms involving derivatives, in (\ref{3.15}) and (\ref{3.16}),
to be equal, we must require that
\be     \label{3.17}
2 \li \qo \qi \kim  \, =\, \qo \lt \kim + \qo \li \ktm
\ee
In that case we can set
\be     \label{3.18}
\qo \ki ^{(\mu}\kt ^{\nu )}\, = \, 2 \qi \kim \kin
\ee
We can thus eliminate $\qi$ from the two index (symmetric)
fields and reduced their number to one.
In (\ref{3.17}), we can observe a pattern:
$\qi$ attaches itself, first to $\lambda _{1}$ converting
it to $\lambda _{2}$, and then to $\ki$ converting it to $\kt$. The same
pattern is evident in (\ref{3.18}).  We will observe this in other
cases also.

Let us turn to the fields with one Lorentz index, of which there are
four: $\qt \kim , \, \qi \ktm ,\, \qi \qi \kim ,\, and \, k_{3}^{\mu}$.
Applying the identifications (\ref{3.17}), we find that:
\be     \label{3.19}
\delta [ \qt \kim ] \, =\, \frac{1}{2} \li \ktm \qo +
\frac{3}{2} \lt \kim \qo + \li \qt \ko
\ee
\be     \label{3.20}
\delta [\qi \ktm ]\, = \, \frac{3}{2} \li \ktm \qo +
\frac{1}{2} \lt \kim \qo + \lt \qi \kom
\ee
\be     \label{3.21}
\delta [\qi \qi \kim ] \,= \, \li \ktm \qo ^{2} + \lt \kim \qo ^{2}
+ \li \qi \qi \ko
\ee
\be     \label{3.22}
\delta [k_{3} \qo ^{2}]\, =\, \li \ktm \qo ^{2} + \lt \kim \qo ^{2} +
\lambda _{3} \qo ^{2} \ko
\ee
We can easily see that
\be     \label{3.23}
\qi \qi \kim \, = \, 1/2(\qt \kim + \qi \ktm )
\ee
is a consistent identification, provided
\be     \label{3.24}
\qi \qi \li \, =\, 1/2 (\li \qt + \lt \qi )
\ee
Note that (\ref{3.23}) and (\ref{3.24}) are also
consistent with the pattern described above.
In any case, we can impose both (\ref{3.23}) and (\ref{3.24}) consistently.
If we further impose
\be     \label{3.25}
\lambda _{3} \qo ^{2} \, =\, \li \qi \qi
\ee
we can also make the identification
\be     \label{3.26}
\qi \qi \kim \, = \, k_{3}^{\mu} \qo ^{2}
\ee

As a result of all these identifications, we have two independent
vectors that we can choose to be $\qt \kim$ and $k_{3}^{\mu}$, and
also two independent vector gauge parameters, $\li \ktm $ and
$\lt \kim$.  These are thus sufficient to gauge the vectors away.

We now turn to the scalars: $\qi \qi \qi , \, \qi \qt , \, q_{3}$.
By comparing their gauge transformations one finds that the identification
\be     \label{3.27}
\qi \qi \qi \, =\, \qt \qi \qo \, = \, q_{3} \qo ^{2}
\ee
is consistent.  The only scalar gauge parameter is $\lambda _{3}$.

One can now rewrite all the free equations without any $\qi$'s.  At
this point one can apply the prescription of the last section to obtain
interacting equations.  Similarly, one can also deduce the non-linear
transformation laws.  We will write down the gauge transformation laws
and one of the equations by way of illustration.

The gauge tranformation laws are:
\[
\delta \phi _{111 \mu \nu \rho} \, =\, (\mup \epsilon _{\nu \rho}
+\Lambda _{2\nu}\mup A_{\rho} +\Lambda _{2\rho}\mup A_{\nu} +
\Lambda _{1}\mup S_{\nu \rho}) +(\mu \leftrightarrow \nu ) +(\mu
\leftrightarrow \rho ) .
\]
\[
\delta (\ki _{(\mu}\kt _{\nu )}) \equiv \delta \phi_{(12)\mu \nu}
\, = \, \{\mup \epsilon _{1\nu } +\mup \epsilon _{2\nu} +
\epsilon _{\mu \nu}
\]
\[
+\Lambda _{2\mu}A_{\nu} +\Lambda _{2} \mup A_{\nu} +
\Lambda _{1} \mup S_{\nu }\} +
\{ \mu \leftrightarrow \nu \} .
\]
\[
\delta \phi _{3\mu} \, = \,\mup \epsilon +  \epsilon _{\mu}
+\epsilon _{2\mu}
+\Lambda _{2} A_{\mu} +\Lambda _{1}S_{\mu}.
\]
\[
\delta (\qt \kim ) \equiv \delta \phi _{12\mu} \, =\,
\mup \epsilon _{2}
+\frac{1}{2}\epsilon _{1\mu}
+\frac{3}{2}\epsilon _{2\mu}  + \frac{1}{2}  \Lambda S_{\mu}
-\frac{1}{2} \Lambda _{2} A_{\mu} +\Lambda \mup \eta .
\]
\be     \label{3.28}
\delta \eta _{3} \, = \, \epsilon
\ee
Where
\be     \label{3.29}
\li \qt \sim \epsilon _{2}\, ;\,
\lambda _{3} \sim \epsilon \, ; \,
\li \ktm \sim \epsilon _{2}^{\mu} \, ; \,
\lt \kim \sim \epsilon _{1}^{\nu} \, ; \,
\li \kim \kin \sim \epsilon ^{\mu \nu }.
\ee
 Finally, there is a tracelessness condition that the gauge parameters
 of the free theory satisfy \cite{bslv,bslv1};
\[
\li \ki .\ki + \li \qi \qi (=\lambda _{3} \qo ^{2}) \, =\, 0,
\]
or
\be     \label{3.30}
\epsilon ^{\nu}_{\nu} +  \epsilon \qo ^{2} \, = \, 0.
\ee
which gets modified to
\be     \label{3.31}
\epsilon ^{\nu}_{\nu} + 2 \Lambda _{2}^{\nu}A_{\nu} +
\Lambda A^{\nu}A_{\nu}
+\epsilon \qo ^{2} \, = \, 0.
\ee
in the interacting theory.

The equation for $k_{3}^{\mu}$ in the free theory is \cite{bslv}
($\qo ^{2} = 4$):
\[
(\ko ^{2} + \qo ^{2}) k_{3}^{\mu} -\ko ^{\mu} (q_{3}\qo + k_{3}.\ko )
-\kim (\qt \qo +\kt .\ko ) -
\]
\be     \label{3.32}
\ktm (\qi qo + \ki .\ko ) + \kim (\ki .\ki +\qi \qi )
+2(\kom (\kt .\ki + \qt \qi ) \, = \, 0.
\ee
First $\qi \ktm$ is replaced by $(2k_{3}^{\mu} \qo - \qt \kim)$,
$\qi \qi \kim $ by $k_{3}^{\mu} \qo ^{2}$ and $\qt \qi$ by $q_{3}\qo$.
Then, following the rule for introducing interactions, we replace $k _{i} $
by $\int d\s k_{i} (\s )$ and $q_{i} $ by $\int d \s q_{i} (\s )$
in (\ref{3.32}):

We have the following relations:
\[
\int d\si d\st \qt (\si ) \kim (\st ) \,= \,
\phi _{12}^{\mu} + \eta A^{\mu},
\]
\[
\int d\si d\st
\ki ^{(\mu}(\si ) \kt ^{\nu )}(\st ) \, =
\, \phi _{(12)}^{\mu \nu} + A^{(\mu}S^{\nu )},
\]
and
\[
\int d\si d\st d\sth
\kim (\si )\kin (\st )\ki ^{\rho}(\sth ) \, =\,
\]
\be     \label{3.33}
\phi _{111}^{\mu \nu \rho } + S^{\mu \nu} A^{\rho} +S^{\mu \rho}A^{\nu}
+S^{\nu \rho} A^{\mu} +A^{\mu}A^{\nu}A^{\rho}
\ee
Inserting (\ref{3.33}) in (\ref{3.32}) we get an interacting equation
that is invariant under the gauge transformation (\ref{3.28}). Similar
calculations can be performed for all the other equations given
in \cite{bslv}.  Thus we can obtain equations for a masssive spin-3
field interacting with massive spin-2 and massless spin -1 fields.

\newpage
\section{Conclusions}
\setcounter{equation}{0}

In this paper we have shown how the loop variable approach of
\cite{bslv,bslv1} that was used there to obtain free equations of
motion, can be generalized to include interactions.  The interactions
are obtained by introducing an additional parameter $\s$, that
broadens the loop to a band.  Each loop in the band stands for
a separate string.  If $\s _{1} = \s _{2}$, then we have the same string,
and otherwise two different strings, but at the same space-time point, i.e.
an interaction.  Thus in this approach a free string breaks
up into a number of lighter strings in a well defined way to give
interactions.  The net result is that one obtains a non-linear
equation describing the propagation of a massive mode and interacting
with all the (strictly) lighter modes.  Furthermore gauge
invariance of this equation, which is essential
for consistency of the theory, follows naturally from the
gauge invariance  of the free equation.
This is reminiscent of
what happens in Yang-Mills theories where interactions are introduced
by gauging global symmetries.

The equations do not contain all the non-linear terms that one expects
from string theory.  This is obvious since we only have interactions
with modes that are lighter.  For e.g. $S^{\mu \nu}$ interacts with
$A^{\mu}$ only, but not with $S^{\mu \nu}$ itself - there are no
self interactions.  Thus it corresponds to a certain well defined
truncation.  To understand the nature of this truncation, we turn to the OPE
of two vertex operators:
\[
:\p _{z}X^{\mu}(z)\e : :\p _{w}X^{\nu}(w) \qe :
\]
\[
=\, (z-w)^{k.q}
:\p _{z}X^{\mu}(z)\p _{w}X^{\nu}(w) e^{ik.X(z)+iq.X(w)}:
\]
\be     \label{4.1}
+:\{ (\frac{iq_{0} ^{\mu} \p_{w}X^{\nu}(w)}{z-w}\, - \,
\frac{i\kon \p _{z}X^{\mu}(z)}
{z-w} + \frac{(\delta ^{\mu \nu} +q_{0}^{\mu} q_{0}^{\nu} )}{(z-w)^{2}})
e^{ik.X(z)+iq.X(w)} \} :
\ee
The first term in the RHS of (\ref{4.1}) can be Taylor expanded in powers
of $(w-z)$ and we get:
\[
:\p _{z}X^{\mu}(z)\p _{w}X^{\nu}(w) e^{ik.X(z)+ip.X(w)}:
\]
\be     \label{4.2}
=\, :\p _{z}X^{\mu}(z)(\p _{z}X^{\nu}(z) + (w-z)\pp X^{\nu}(z) +...)
e^{ikX(z)+p[X(z)+(w-z)\partial _{z}X +...]} :
\ee
\be     \label{4.3}
=\, :\p _{z}X^{\mu}\p _{z}X^{\nu} (z)e^{i(k+p)X(z)}: \, +\, O(z-w)
\ee
The presence of a cubic interaction term $S^{\mu \nu}A_{\mu} A_{\nu}$
can be inferred from the first term of (\ref{4.3}).  The next term
would correspond to $\phi _{12}^{\mu \nu} A_{\mu}A_{\nu}$.
However, such a term is not present in the equation of motion of
$\phi_{12}^{\mu \nu}$ derived by the procedure of Sec II.
Similarly,
the second term in (\ref{4.1}) contains, among other things,
a Yang-Mills type term $\mup A_{\nu} [A^{\mu},A^{\nu}]$ (which would
vanish in the U(1) case).  This is also absent from the equations
of motion obtained by the procedure of this paper.

Thus it is clear that we have retained only one
term from the OPE - this is the first term in (\ref{4.3}). $ S^{\mu \nu}$
has the same value (N=2) of the oscillator number, $N=\sum _{n} a^{\dag}_{n}
a_{n}$, as the two states involved in the product, $A^{\mu}$,and $A^{\nu}$.
This is true of all the equations we have written down, i.e. all
the terms of an equation have the same oscillator number.
In the case of the spin-3
field $\phi_{111}^{\mu \nu \rho}$ also, which has $N=3$, one can
check that each term in the equation of motion has $N=3$.  From this
observation we can conclude that the equation of motion that one obtains
by the procedure described in this paper
is a consistent truncation of the full equations of string theory -
a truncation, where
oscillator number is conserved. The consistency reflects itself
in the fact that the equations have the full gauge invariance necessary
for consistent propagation of massive higher spin particles.

What is perhaps most intriguing is the form of
the equations and the gauge transformations.
The interacting equations look exactly like the free equations of a
loop variable that has been broadened to a band.  This is also
true of the gauge transformations.  One can speculate that this band
represents a membrane and that, in some approximation, an interacting
string looks like a {\em free} membrane.  This is somewhat analogous to
large-N Yang-Mills theory being a string theory.

Finally, it should be possible to generalize this construction to
include, in the equations, interaction terms that do not have the same
oscillator number, and thereby obtain the full string equation.
We hope to return to this issue.

\end{document}